# Cloud Migration: A Case Study of Migrating an Enterprise IT System to IaaS

Ali Khajeh-Hosseini       David Greenwood       Ian Sommerville

*Cloud Computing Co-laboratory*
*School of Computer Science*
*University of St Andrews, UK*
*{akh, dsg22, ifs}@cs.st-andrews.ac.uk*

## Abstract

*This case study illustrates the potential benefits and risks associated with the migration of an IT system in the oil & gas industry from an in-house data center to Amazon EC2 from a broad variety of stakeholder perspectives across the enterprise, thus transcending the typical, yet narrow, financial and technical analysis offered by providers. Our results show that the system infrastructure in the case study would have cost 37% less over 5 years on EC2, and using cloud computing could have potentially eliminated 21% of the support calls for this system. These findings seem significant enough to call for a migration of the system to the cloud but our stakeholder impact analysis revealed that there are significant risks associated with this. Whilst the benefits of using the cloud are attractive, we argue that it is important that enterprise decision-makers consider the overall organizational implications of the changes brought about with cloud computing to avoid implementing local optimizations at the cost of organization-wide performance.*

## 1. Introduction

Over the last few years startup companies such as Twitter (www.twitter.com) and Animoto (www.animoto.com) have used clouds to build highly scalable systems. However, cloud computing is not just for startups; enterprises are attracted to cloud-based services as cloud providers market their services as being superior to in-house data centers in terms of financial and technical dimensions e.g. more cost effective, equally or perhaps more reliable, and highly scalable [1, 2, 3, 4]. Whilst the technological and financial benefits may be seductive, it is important that enterprise decision-makers factor in other dimensions, the overall organizational implications of the change, to avoid ignoring other significant factors and thus implementing local optimizations at the cost of organization-wide performance.

There are currently few case studies that investigate the migration of existing IT systems to the cloud [5]. Furthermore, little has been published about the implications of cloud computing from an enterprise or organizational perspective [6].

This paper's original contribution is to address these issues by presenting the results of a case study that investigated the migration of an IT system from a company's in-house data center to Amazon EC2 (www.aws.amazon.com). The primary focus of the case study was on the financial and socio-technical enterprise issues that decision-makers should consider during the migration of IT systems to the cloud.

This case study identifies the potential benefits and risks associated with the migration of the studied system from the perspectives of: project managers, technical managers, support managers, support staff, and business development staff. The paper is based upon data collected from an IT solutions company considering the migration of one of their systems to Amazon EC2. This Infrastructure-as-a-Service (IaaS) layer of the cloud is arguably the most accessible to enterprise as they could potentially migrate their systems to the cloud without having to change their applications. In addition, Amazon Machine Images are readily available for enterprise applications such as Oracle Database and Citrix XenApp (www.aws.amazon.com/solutions/solution-providers/).

The paper is structured such that: the background section introduces the proposed migration project and looks at related work in this area; the method section describes the approach used to collect and analyze data; the results section identifies the cost saving benefits of using cloud computing and its affect on the support and maintenance of the system under investigation; the organizational benefits and risks of the migration are also discussed in the results section; the paper concludes by discussing the main points revealed by this case study and looking at future work.

## 2. Background

### 2.1. Proposed Migration Project

The case study organization is a UK based SME that provides bespoke IT solutions for the Oil & Gas industry. It comprises of around 30 employees with offices in the UK and the Middle East. It has an organizational structure based on functional divisions: Administration; Engineering; Support; of which Engineering is the largest department.

The migration use-case comprises the feasibility of the migration of one of the organization's primary service offerings (a quality monitoring and data acquisition system) to Amazon EC2. The following is an anonymized description of the situation: *Company C* is a small oil and gas company who owns some offshore assets in the North Sea oilfields. Company C needed a data acquisition system to allow them to manage their offshore operations by monitoring data from their assets on a minute by minute basis. Company C's assets rely on the production facilities of *Company A* (a major oil company), therefore the data comes onshore via Company A's communication links. Company C does not have the capabilities to develop their own IT systems, hence they outsourced the development and management of the system to *Company B*, which is an IT solutions company with a small data center. Figure 1 provides an overview of the system, which consists of two servers:
1. A database server that logs and archives the data coming in from offshore into a database. A tape drive is used to take daily backups of the database, the tapes are stored off-site.
2. An application server that hosts a number of data reporting and monitoring applications. The end users at Company C access these applications using a remote desktop client over the internet.

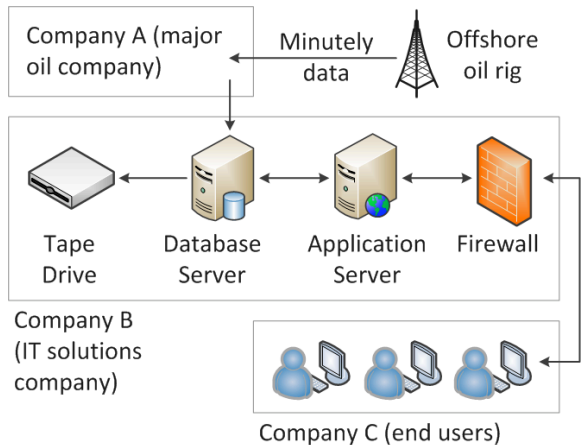

**Figure 1: System Overview**

The system infrastructure was deployed in Company B's data center and went live in 2005. Since then, Company B's support department have been maintaining the system and solving any problems that have risen. This case study investigated how the same system could be deployed using the cloud offerings of Amazon Web Services. Figure 2 provides an overview of this scenario, where Company B deploys and maintains the same system in the cloud.

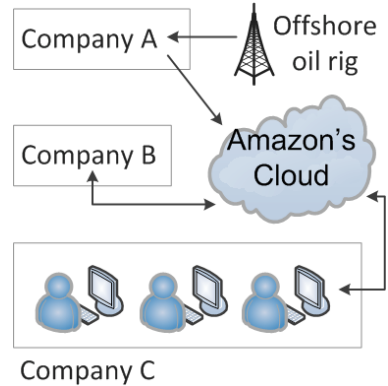

**Figure 2: System deployed in the cloud**

## 2.2. Related Work

Cloud computing is not just about a technological improvement in data centers; it represents a fundamental change in how IT is provisioned and used [7]. For enterprises to use cloud computing, they have to consider the benefits, risks and effects of cloud computing on their organizations. Case studies provide an effective way to investigate these areas in real-life organizations. This section takes a brief look at the related work in each of these three areas.

Armbrust *et al.* [1] argued that elasticity is an important economic benefit of cloud computing as it transfers the costs of resource over-provisioning and the risks of under-provisioning to cloud providers. Motahari-Nezhad *et al.* [8] added that the potentially reduced operational and maintenance costs is also important from a business perspective. Walker [9] also looked into the economics of cloud computing, and pointed out that lease-or-buy decisions have been researched in economics for more than 40 years. Walker used this insight to develop a model for comparing the cost of a CPU hour when it is purchased as part of a server cluster, with when it is leased (e.g. from Amazon EC2). Walker's model was a good first step in developing models to aid decision makers, but it was too narrow in scope as it focused only on the cost of a CPU hour.

Klems *et al.* [10] presented as a framework that could be used to compare the costs of using cloud computing with more conventional approaches, such as using in-house IT infrastructure. Their framework was very briefly evaluated using two case studies. However, no results were provided because the framework was at an early developmental stage and more conceptual than concrete. In contrast, we provide detailed results by comparing the costs of using an in-house data center with AWS for our case study.

From an enterprise perspective, security, legal and privacy issues seem to present a number of risks as pointed out by detailed reports from the Cloud Security Alliance [11] and European Network and Information

Security Agency [12]. Others have discussed risks posed by a cloud's geographic location [13], legal issues that affect UK-based organisations [14], and the technical security risks of using cloud computing [15].

However, not much has been published about the organizational risks of the change that cloud computing brings to enterprise. Yanosky [16] discussed how cloud computing will affect the authority of the IT department within universities and argued that the IT department's role will change from "provider to certifier, consultant and arbitrator". This could lead to inefficiencies in organizations if certain stakeholders resist the changes brought about by cloud computing. One approach to understanding these risks is to capture each stakeholders' perception of the change through semi-structured interviews allowing stakeholders to raise the benefits, risks, opportunities or concerns as they perceive them [17, 18].

The results of the case study presented in this paper are novel as they attempt to highlight the overall organizational implications of using cloud computing. This issue has not been discussed to any significant extent in the previously mentioned literature.

## 3. Methodology

This case study involved fieldwork at Company B's offices between May to July 2009. Initially, all documents relating to the system under investigation were gathered and studied. The fieldwork had three stages:

**Stage 1:** The infrastructure costs of the system were calculated from project reports and invoices. These costs were compared with the costs of a similar infrastructure setup on Amazon EC2.

**Stage 2:** Company B has a database of all support and maintenance issues regarding the systems that they support. This database was manually researched and all of the support calls that would potentially be affected by the migration were identified and analyzed.

**Stage 3:** The results from the above two stages were used to produce a poster. The poster was presented to Company B's employees and six semi-structured interviews were performed at their offices. The interviews started by giving the interviewees an overview of Amazon EC2 as they were only partially familiar with this technology. Each interview was recorded and a transcript of each interview was produced. Each transcript was read be two researchers (one present at the interview and one not) and a number of issues were identified and agreed using a stakeholder impact analysis. Stakeholder impact analysis is a method of identifying potential sources of benefits and risks from the perspectives of multiple stakeholders, and is performed by analyzing interview transcripts. It comprises of:

1. Identifying key stakeholders;
2. Identifying changes in what tasks they would be required to perform and how they were to perform them;
3. Identifying what the likely consequences of the changes are with regards to stakeholders time, resources, capabilities, values, status and satisfaction;
4. Analyzing these changes within the wider context of relational factors such as tense relationships between individuals or groups to which stakeholders belong;
5. Determining whether the stakeholder will perceive the change as unjust (either procedurally or distributively) based upon changes and their relational context.

The results of the case study fieldwork are discussed in the next section.

## 4. Results

### 4.1. Infrastructure Costs

Company C paid £104,000 to Company B for the system in 2005, £19,400 of which was for the system's infrastructure; the rest of the costs were for system development and deployment. The infrastructure included two servers (each having two Intel Xeon 3.4GHz processors, 2GB RAM, 6 x 72GB hard drives in a RAID 10 array resulting in around 200GB of effective storage, Windows Server 2003 OS), a tape drive, network equipment, a server rack, shelf spares. In addition, Company C pays £43,000 per year to Company B for system support and maintenance, £3,600 of which is for the running costs of the system infrastructure. Over a five year period, the total cost of the system infrastructure is therefore: £19,400 + (5 x £3,600) = £37,400. We acknowledge that hardware performance has changed since 2005 and perhaps it may be perceived that costs should have reduced, however in reality they remain similar. For example, the servers used by Company C cost £4,525 in 2005, the ones used in a similar project in 2009 cost £4,445.

Amazon EC2 provides an option of using either small or large server instances depending on the amount of CPU power and RAM required. The system could initially run on two small instances as the application and database server do not seem to be under a heavy load. However, this could be changed for large instances if the performance is found to be unacceptable. This would not have been possible using the existing approach since all hardware must be purchased before the system is deployed, and cannot easily be changed afterwards.

Table 1 shows a comparison of the costs of the system infrastructure, the amounts have been rounded to the nearest £10. The following specifications were used to calculate the costs of running the system on AWS: two Microsoft Windows On-Demand instance (AWS do not offer reserved instances for Windows) in Europe running 730 hours per month (i.e. 24x7); 20GB data transfer in; 20GB data transfer out; 200GB EBS storage (i.e. amount of effective storage on existing servers), 100 million EBS I/O request; 30GB EBS snapshot storage (for daily backups); 10 snapshot GET requests (in case backups need to be retrieved); 30 snapshot PUT requests (for daily backups).

**Table 1: Comparison of infrastructure costs between cloud and Company B's data center**

| Period | Amazon Server Instances | | | Cmpny B |
|---|---|---|---|---|
| | 2 small | 1 small + 1 large | 2 large | |
| 1 Month | £200 | £390 | £590 | £620 |
| 1 Year | £2,400 | £4,680 | £7,080 | £7,440 |
| 5 Years | £12,000 | £23,400 | £35,400 | £37,200 |

From Company B's perspective, the cloud presents an opportunity to bid for new projects without having to worry about space in their data center as they are currently running out of rack space, and building a new data centre is an expensive venture. It also means that they could propose a cheaper alternative to deploying systems in their in-house data center for their clients.

From Company C's perspective (the end users), Table 1 shows that the cost of running their system in the cloud is cheaper than using Company B's data center. For example, it would be 37% cheaper to deploy the system in the cloud assuming that a small and a large server instance are used. Furthermore, no upfront capital is required for infrastructure in the cloud since users are charged on a monthly basis. The potential cost reductions certainly seem significant, but the affects of a migration on the support and maintenance of the systems must also be considered.

### 4.2. Support and Maintenance

The system is currently supported and maintained by Company B's support department who also perform regular health checks to ensure that the system is running as expected. The health checks involve checking error logs, backup logs, server load levels, communication links etc. The support and maintenance of the system would be affected if the system was migrated to the cloud since the support department would no longer have full control over the system infrastructure.

Company B maintains a database of all the support calls they receive by telephone or email either externally from end users or internally from support engineers doing regular heath checks. Since the system went live in 2005, 218 support calls have been made regarding the operation of the system. The majority of these calls were about software problems, however, the titles of all calls were studied and a shortlist of 112 calls was made for further investigation. It was found that the following 45 calls were related to the system's infrastructure:

- 38 calls were related to backup problems between the database server and the tape drive. Common problems included faulty tapes, failed backup attempts, and even loose cables presumably related to tapes being taken in and out of the drive on a daily basis. These problems were usually fixed by erasing the tapes, rebooting the tape drive or re-running backup scripts, but there were a few occasions when no backup was taken for that day.
- 5 calls were related to network problems, one of which required a router to be rebooted, and another that was caused by a power cable being unplugged accidentally.
- 2 calls were related to power outages at Company B's data center.

The previously mentioned calls could potentially have been eliminated if the system was deployed in the cloud since Amazon would be responsible for hardware related issues. As shown in Figure 3, this accounts for around 21% of the support calls but it should be noted that some additional calls might be introduced if the system was migrated to the cloud. These cloud related issues could include power outages at Amazon's data centers or network latency issues; however, the important point is that these issues would be dealt with by Amazon. This could be seen as a big advantage for Company B's support department as it allows them to focus on software related issues, which are more important to the end users.

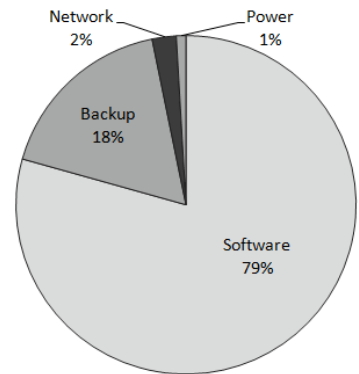

**Figure 3: Overview of support calls**

### 4.3. Stakeholder Impact Analysis

Analysis of the interview data suggests that the proposed cloud migration would have a positive net benefit from the perspective of the business development functions of the enterprise and the more junior levels of the IT support functions. A perceived zero net benefit was perceived by the project management and support management functions of the enterprise. A negative net benefit was perceived by the technical manager and the support engineer functions of the enterprise.

Stakeholder impact analysis data suggests that there are numerous potential benefits but also risks associated with the migration of the system to the cloud. Tables 2 and 3 summarize the benefits and risks of the migration as identified by the stakeholder impact analysis. The second column in Tables 2 and 3 refers to the number of specific benefits/risks identified, and hence indicates the distribution of benefit or risk across different areas. Twelve specific benefits were identified in contrast to eighteen specific risks. According to the analysis the largest source of benefit to be derived from the cloud providing an opportunity to manage income and outgoings in a new way, followed by the opportunities to offer new products/service, improved job status and removal of tedious work. The largest source of risk will be derived from the potential deterioration of 'customer care and service quality', 'increased dependence on $3^{rd}$ party', decrease in satisfying work and increases of workload.

**Table 2: Sources of benefit identified by stakeholder impact analysis**

| Benefits | # |
|---|---|
| Opportunity to manage income & outgoings | 3 |
| Opportunity to offer new products/services | 2 |
| Improved status | 2 |
| Removal of tedious work | 2 |
| Improve satisfaction of work | 1 |
| Opportunity to develop new skills | 1 |
| Opportunity for organizational growth | 1 |

**Table 3: Sources of risk identified by stakeholder impact analysis**

| Risks | # |
|---|---|
| Deterioration of customer care & service quality | 3 |
| Increased dependence on external $3^{rd}$ party | 3 |
| Decrease of satisfying work | 3 |
| Departmental downsizing | 2 |
| Uncertainty with new technology | 2 |
| Lack of supporting resources | 1 |
| Lack of understanding of the cloud | 1 |

## 4.4. Benefits

### 4.4.1. Opportunity to manage income & outgoings.

Introducing third party cloud infrastructure solutions presents itself as an opportunity to improve the management of income and outgoings for both finance staff and customers. Third party cloud infrastructure solutions facilitate the easing of cash-flow management for finance staff as the cloud pricing model has minimal upfront cost and monthly billing, and it also minimizes variability of expenditure on electricity. These are a benefit, in contrast to in-house data center, as upfront costs of buying hardware are high and clients can be slow to pay, resulting in cash-flow difficulties. Additionally energy costs are a significant outgoing and by using an external provider they would benefit from providers ability to negotiate whole-sale energy prices.

Third party cloud infrastructure solutions also surface many opportunities for managing income for customers, sales and marketing staff, as new pricing models can be offered to them. This is a benefit, in contrast to internal data centers which require a pricing to model comprising of a large upfront fee plus monthly support costs (due to cash-flow issues), as customers can be offered more choice over how they want pay or alternatively the finance department can choose to get the infra-structure outsourcer to bill their customers directly reducing the finance departments' administrative burden.

### 4.4.2. Improved status.
Introducing third party cloud infrastructure solutions present an opportunity for support management and support engineers to improve their status. Support managers can improve their status in the organisation by successfully championing the high profile migration that has strategic implications. This is a benefit to the support manager as by working with new and potentially prestigious technology it may lead to career progression and increased job satisfaction. Support engineers would also benefit by improving their status within their industry by developing sought after cloud administration skills and experience.

### 4.4.3. Improve satisfaction of work.
Third party cloud infrastructure solutions present an opportunity for support engineers, sales and marketing staff to improve the satisfaction of their work. It is an opportunity for support engineers to shed unsatisfying routine and potentially time consuming work such as performance of hardware support, network support and switching backup tapes as well as being offered new challenges in terms of cloud administration. This is a benefit as support engineers can focus on more satisfying and value-adding work such as resolving customers' software support requests. This benefit is enabled by the switch to cloud-infrastructure as the third party cloud provider would be responsible for the more routine maintenance.

Technical developers could also benefit from the migration as they can be involved in systems support (e.g. performing regular system health checks), which are sometimes viewed as a chore. In small organizations, there is not usually a clear distinction between the roles of system administrators and technical developers, and different people have to be involved when there is a problem.

Third party cloud infrastructure solutions present an opportunity for sales and marketing staff to create new product/service offerings that better fit the customers need in terms of scalability and cost effectiveness in contrast to an in-house data center. This is a benefit as this provides staff with a new and potentially satisfying challenges that would not have existed without the migration to cloud-infrastructure.

**4.4.4. Opportunity to develop new skills.** Third party cloud infrastructure solutions present an opportunity for support managers, engineers, sales and marketing staff to develop new skills. For support managers and engineers it is an opportunity to develop new skills in cloud computing administration. This is a benefit as the support engineers will expand their existing skill sets and experience with knowledge of managing a technology that will be in demand throughout the IT industry for years to come. For sales and marketing staff it presents an opportunity to develop skills is product/service creation and launching. This is a benefit to sales and marketing staff as it will expand their existing skill sets and experience enabling their career progression.

**4.4.5. Opportunity for organizational growth.** Third party cloud infrastructure solution presents an opportunity for sales and marketing staff to create new product/service offerings that may appeal to a larger market-share due to cloud-infrastructures properties of scalability and its cost effectiveness in contrast to an in-house data center. This is a benefit as it may facilitate sales staff meeting targets by enabling them to target market segments previously not attracted by limitations of scalability.

## 4.5. Risks

**4.5.1. Deterioration of customer care & service quality.** Third party cloud infrastructure solutions present a risk to customer care and overall service quality for support managers, support engineers and customer care staff. Support managers and engineers are at risk of becoming dependent upon a cloud service provider which they have no control over and at risk of requiring additional resources to do the migration and deal with short term issues that arise subsequent to the migration (e.g. shortfalls in cloud operations knowledge resulting in tasks taking temporarily longer to complete). Support managers and engineers specifically risk becoming dependent upon a cloud service provider for resolving hardware and network issues. This is a risk as it could result in the deterioration of service quality that the support manager would not be able to control. Support managers also risk temporarily requiring more resources to cope with migration and also the relative lack of knowledge and experience held by support staff regarding cloud systems. This is a risk because staff may initially require more time to perform the same tasks due to having to learn how-to perform tasks in the cloud environment which could compromise service quality and customer service.

Customer care staff are also at risk of not being able to offer the existing levels of customer service as it may take longer to resolve customer queries as cooperation with external service providers may become necessary. This is a risk because response times to deal with customer queries may increase resulting in back-logs and cascades of additional work as customer call back for progress updates and will result in customer care staff dissatisfaction.

**4.5.2. Decrease in satisfaction.** Third party cloud infrastructure implementations present a risk of decreasing job satisfaction of support engineers, sales & marketing staff, and customer care staff. Support engineers risk decreasing job satisfaction as work may shift from a hands-on technical role to reporting and chasing up issues with third party service providers. Support engineers will become dependent upon the responsiveness of third party service providers to resolve problems thus reducing the level of control support engineers have over resolving issues.

This is a risk to support engineer satisfaction as they derive satisfaction from technical aspects of work and rapidly resolving problems to customer satisfaction. Sales and marketing staff risk of decreasing job satisfaction if they are set unrealistic goals regarding the selling of the new cloud based services. This is a risk to sales and marketing's satisfaction as they derive satisfaction from meeting sales and market share targets. Customer care staff also risk decreasing job satisfaction because their ability to perform their job will be dependent upon third parties out of their control resulting in a greater lag between customer queries and resolution.

**4.5.3. Departmental downsizing.** Third party cloud infrastructure implementations present a risk of downsizing to IT support departments. IT support departments are at risk of downsizing if the majority of their work comprises hardware and network support. This is a risk because cloud providers will be responsible for maintaining these aspects of support making the capability unnecessary within the IT support department. Both support managers and support engineers will be impacted as support engineers may lose their jobs and the support managers may lose influence as they have a small department.

**4.5.4. Uncertainty with new technology.** Third party cloud infrastructure implementations present a risk to the finance/business development staff as it may open the organization to long-term volatility derived from market forces associated with the costs of using a cloud and data transfer costs. This is a risk as the medium to long-term viability of a cloud solution versus an internal hosting solution are uncertain. Additionally, switching to external hosting decreases the certainty of customer lock-in in terms of software support contracts as now the hardware is maintained externally and therefore the company can no longer make the case that it offers an 'all-in-one' maintenance contract which avoids having to deal with multiple contactors. Another consideration is the loss of in-house expertise resulting in additional barriers to bringing the system back in-house if the cloud provider is inadequate.

**4.5.5. Lack of supporting resources.** Third party cloud infrastructure implementations present a risk resource scarcity in IT support and sales/marketing departments. There is a risk of temporarily upsizing the IT support departments to cope with migration and also the relative lack of knowledge and experience held my support engineers regarding cloud systems. This is a risk because staff may initially require more time to perform the same tasks due to having to learn how-to do so in the cloud environment. There is a risk of temporarily upsizing sales/marketing to cope with the creation and launch of new cloud based products/services. This is a risk because sales and marketing staff will need to develop appropriate strategies and materials to ensure the marketplace is aware of the product offering.

In summary these results illustrate that whilst the financial and technological analyses are certainly important, the organizational dimension should also be considered. This should be particularly considered from service quality and customer care perspective, and the organizational governance and risk implications of being so highly dependent upon a third party for product/service delivery to customers. In some cases, the financial dimension may not even be the primary consideration for business-critical applications. These findings are reinforced by the fact that at present the majority of management at the organization is reluctant to implement the change beyond a test environment despite the financial incentives as the risks are perceived to outweigh the lost savings.

## 5. Conclusion and Future Work

Cloud computing is a disruptive technology that is set to change how IT systems are deployed because of its apparently cheap, simple and scalable nature. The findings of this case study show that cloud computing can be a significantly cheaper alternative to purchasing and maintaining system infrastructure in-house. Furthermore, cloud computing could potentially eliminate many support-related issues since there would be no physical infrastructure to maintain. Despite these advantages, this case study showed that there are important socio-technical issues that need to be considered before organisations could migrate their IT systems to the cloud.

The system infrastructure in the case study would have cost 37% less over 5 years on Amazon EC2, and using cloud computing could have potentially eliminated 21% of the support calls for this system. These findings seem significant enough to call for a migration of the system to the cloud but our stakeholder impact analysis revealed that there are significant disadvantages tied to the promised benefits. The disadvantages include risks to customer satisfaction and overall service quality due to diffusion of control to third parties; decreased job satisfaction due to changes in nature of work; and opening the organisation to long term cost volatility in terms of cloud-usage and data transfer costs.

This case study has practical implications for industrial practitioners assessing the benefits of external cloud infrastructures for their organisation. The generic benefits identified can be leveraged to gain buy-in from stakeholders whilst the generic risks identified should be adapted into a risk register and monitored to ensure their projects do not fall prey to common cloud infrastructure migration risks. Additionally the stakeholder impact analysis method may be adopted by enterprises and performed in-house to create a bespoke understanding of their situation.

The limitations of this study are that the cost analysis only focused on system infrastructure costs, and did not quantify: the cost of doing the actual migration work; how the support staff costs would be affected by the migration; the cost of a support contract that might be required with AWS Premium Support (www.aws.amazon.com/premiumsupport). Support staff costs are difficult to quantify as they would first require the system to be migrated to the cloud and run for a period of time to study any issues that would arise. There are also longer-term costs associated with the migration of systems, such as the cost of migrating to another cloud provider if the current provider is inadequate or raises their cost plans, or even the costs associated with the loss of experience/knowledge if a company needs to re-deploy the system in-house. However these limitations are addressed qualitatively by the stakeholder impact analysis which identifies a range of non-quantifiables and evaluates their consequences.

An open issue for future research is that whilst it is clear that from a financial perspective, end-users could benefit from cloud computing it is unclear whether it will materialize for the majority, as many organizations outsource their IT to system integrators and it is at present unclear whether there is sufficient financial incentive for these system integrators. On first impression it may appear that system integrators profits will not rise (assuming IaaS costs are passed directly to the end-user) and may even be marginally less due to loss of small profits associated with hardware sales. However this ignores the following facts: i) that the system integrator will be able to focus their resources and effort on performing value-adding and more profitable activities (e.g. system software support) rather than hardware builds and hardware maintenance; ii) the system integrator will not be paying in-house hosting costs e.g. electricity, cooling, off-site tape archiving. Both facts indicate that in the medium to long term the integrator will be more profitable per unit of work performed. These arguments however require empirical substantiation.

As part of future work, we are working with Company C (the IT solutions company) who are currently considering the migration of non business-critical applications such as those used in their training courses to the cloud. We will be developing decision support tools that can be used to estimate the costs of running IT systems in the cloud, and will continue to investigate the associated risks of doing so. We are also planning to use other case studies to confirm the results presented in this paper, and further develop and evaluate our stakeholder impact analysis method.

## 6. Acknowledgements

We thank the Scottish Informatics and Computer Science Alliance (SICSA) and the EPSRC for funding the authors. We also thank our colleagues at the UK's Large-Scale Complex IT Systems Initiative (www.lscits.org), especially Ilango Sriram, for their comments.